\documentclass[aps,prx,showpacs,twocolumn,amsmath,amssymb,superscriptaddress,floatfix]{revtex4}

\usepackage{epsfig}
\usepackage{epstopdf}
\usepackage{graphicx}
\usepackage{grffile}
\usepackage{bm}
\usepackage{sidecap}
\usepackage{mathptmx}
\usepackage{txfonts}
\usepackage[colorlinks,citecolor=blue,linkcolor=Fuchsia]{hyperref}
\usepackage[usenames,dvipsnames,svgnames]{xcolor}
\usepackage{array}
\usepackage{float}

\usepackage{lipsum}

\newcommand{\tobedeleted}[1]{\textcolor{green}{#1}}
\renewcommand{\tobedeleted}[1]{\relax}

\begin{document}
\renewcommand{\thefigure}{\arabic{figure}}
\def\be{\begin{equation}}
\def\ee{\end{equation}}
\def\ber{\begin{eqnarray}}
\def\eer{\end{eqnarray}}

\def\kv{{\bf k}}
\def\bfr{{\bf r}}
\def\qv{{\bf q}}
\def\pv{{\bf p}}
\def\sigmav{{\bf \sigma}}
\def\tauv{{\bf \tau}}
\newcommand{\h}[1]{{\hat {#1}}}
\newcommand{\hdg}[1]{{\hat {#1}^\dagger}}
\newcommand{\bra}[1]{\left\langle{#1}\right|}
\newcommand{\ket}[1]{\left|{#1}\right\rangle}

\title{Impurity scattering on the surface of topological insulator thin films}
\date{\today}

\author {Mahroo Shiranzaei}
\affiliation{School of Physics, Damghan University, P.O. Box 36716-41167, Damghan, Iran}

\author{Fariborz Parhizgar}\email{fariborz.parhizgar@ipm.ir}
\affiliation{School of Physics, Institute for Research in
Fundamental Sciences (IPM), Tehran 19395-5531, Iran}

\author{Jonas Fransson}
\affiliation{Department of Physics and Astronomy, Uppsala University, Box 516, SE-751 21, Uppsala, Sweden}

\author{Hosein Cheraghchi}\email{cheraghchi@du.ac.ir}
\affiliation{School of Physics, Damghan University, P.O. Box 36716-41167, Damghan, Iran}

\begin{abstract}
We address the electronic structure of the surface states of topological insulator thin films with embedded local non-magnetic and magnetic impurities. Using the $T$-matrix expansion of the real space Green's function, we derive the local density of electrons states and corresponding spin resolved densities. We show that the effects of the impurities can be tuned by applying an electric field between the surface layers. The emerging magnetic states are expected to play an important role both in ferromagnetic mechanism of magnetic topological insulators as well as in its transport properties. In the case of magnetic impurities, we have categorized the possible cases for different spin-directions of the impurities as well as the spin-direction in which the spin resolved density of electron states is calculated and related this to the spin susceptibility of the system.
\end{abstract}
 \maketitle

\section{Introduction}
Topological insulators (TIs), a new state of materials with gapped bulk states and symmetry-protected gapless edge states have recently attracted a great deal of attention in theoretical and experimental studies \cite{prl95Q, prl96, prl95Z, prl97, science314, science318, nature438}. Historically, these gapless edge states, which arise from band inversion, were discovered firstly in two-dimensional (2D) TIs based on HgTe quantum wells \cite{science314, science318}. Only later, angle-resolved photoemission spectroscopy (ARPES) analysis of bismuth-based materials \cite{prl983D, prb76fu, nature452, nature398, science325} revealed the presence of a single Dirac cone at the $\Gamma$ point in the spectrum of the surface states, thus, demonstrating the predictions also for three-dimensional (3D) TIs. These types of 3D TIs such as Bi$_2$Se$_3$, consist of layers interacting via Van-der-Waals \cite{nature438} interaction. Each layer, known as quintuple layer (QL), consists of Bi and Se atoms located in five surfaces so that for thicknesses above 6QLs, the bismuth-based material becomes TI with gapless surface states \cite{nature584}.

The gapless surface states are topologically protected by time-reversal symmetry which forbids backscattering from non-magnetic impurities. In fact, owing to strong spin-orbit interaction in 3D TIs, spin-momentum locking of the surface states is the main theoretical reason for backscattering off impurities that break time-reversal symmetry \cite{nature438,Sessi-2014}.

In 3D TIs, there are numerous studies which have tried to shed light on the effect of impurities \cite{science329659, prl106, prl102, Jonas, weak-antilocalization,Lee-2013, coulomb, Bauer-transport, Sessi-2014, Jiang2015, peixoto}. However, because of some experimental contradictory results, scattering by magnetic and non-magnetic impurities deposited on the surface of TIs is still a controversial topic and has a non-clear picture. There are several explanations for describing enhanced experimental backscattering arising from different types of non-magnetic impurities deposited on the surface of 3D TIs \cite{Bauer-transport, Sessi-2014}. This enhanced backscattering is addressed to spatial distribution \cite{Jonas} and concentration of impurities and defects, and also electron scattering by the step edges in the surface \cite{Bauer-transport} of TI thin films.

On the other hand, bulk-doped magnetic impurities in 3D TI give rise to a local gap around the Dirac cone \cite{science329659, prl106, prl102} which is attributed to ferro-magnetically ordered impurities leading to a magnetic field-induced gap \cite{magnetic_gap1, magnetic_gap2}. However, the gap opening in the Dirac cone cannot emerge at the stage of a single impurity while its local density of states (LDOS) as well as spin-LDOS will be remarkably changed \cite{balatsky}. Moreover, it was shown that the scattering and also transport properties of TIs depend on the polarization direction of the magnetic impurity \cite{Amir,Sessi-2014}.
The effect of Coulomb magnetic scatterers \cite{coulomb} and weak localization in the presence of non-magnetic impurities were the subject of other researches too \cite{weak-antilocalization}.

\begin{figure}[t]
\begin{center}
\includegraphics[width=0.95\columnwidth]{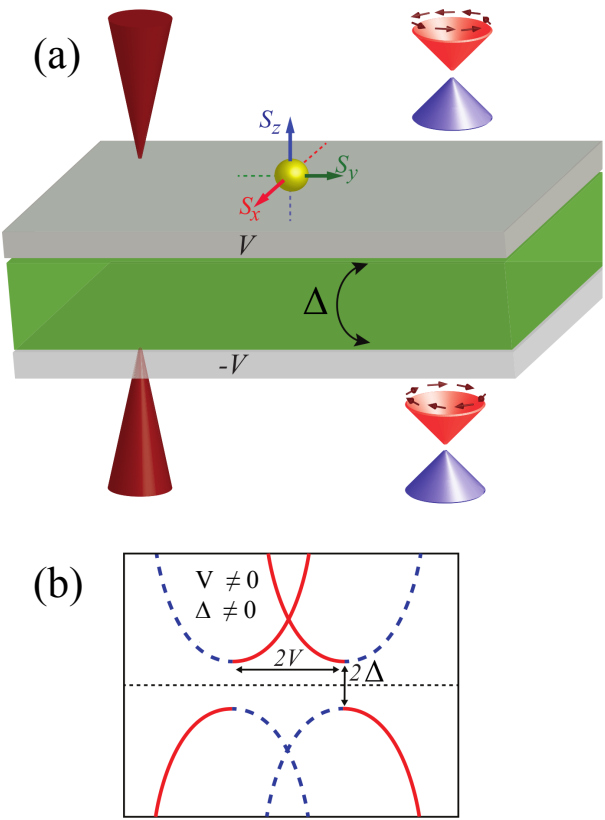}
\end{center}
\caption{
(Color online) (a) Schematic of a TI thin film (green area) with its surfaces (gray area) with opposite helical modes for different surfaces. Here, $ V $ shows applied potential and $ \Delta $ is hybridization between different surfaces. Red cones on the upper and lower the surfaces illustrate STM tips that show the measurement can be done on both surfaces. A single impurity is illustrated as a yellow sphere which can be non-magnetic or magnetic impurity. In the case of magnetic, it can be aligned in one of the spin directions $ \hat{x}, \hat{y}, \hat{z} $. (b) Schematic of band energy dispersion for the TI thin film, Eq. \eqref{eq:3}. Here the blue dashed lines (the red solid lines) show the dispersion coming mostly from the upper (lower) surface.
}
\label{fig:1}
\end{figure}

Although itinerant electrons in the surface of TIs are supposed to provide dissipation-less transport due to the lack of back-scattering, in practice, the bulk states also play a role in the transport properties \cite{J-Zhang2015} and make realization of the pure surface states hard. This problem also occurs when scattering by surface impurities is disturbed by those scatterings coming from the bulk doping \cite{science325}. One way to reduce the effect of bulk states is to use thin slabs  \cite{Hoefer-2014}. For TIs with thickness 5QLs and thinner, the two surface states hybridize with each other giving rise to a gap in the surface state's energy dispersion \cite{nature584}.
Such ultra-thin TIs can be considered as double layers Rashba materials which according to their additional degree of freedom can provide interesting features \cite{Hamiltonian, parhizgar-opt, parhizgar-SCDL}. Also, the bilayer-like material opens the possibility to engineer the band structure of TI thin film by applying an electric field perpendicular to the surfaces. Such field (such phenomenon can originate from the effect of substrate) would separate the degenerate band dispersion like a Rashba splitting as depicted in the Fig. \ref{fig:1}(b). This extra electrical tunability of TI thin films unveils them very favorable to be used in topological magnetoelectric technology and also spintronic devices \cite{effective, rmp76, rmp80, nature378, rmp86}. It was theoretically predicted that a topological transition from 2D quantum spin Hall (QSH) states to normal insulator can be induced by an applied electric field $(V)$ after a critical value \cite{prb81041307, prb80, electrically, effective}. Also it has been shown that special form of tunneling between different surfaces together with possibility of Rashba-type splitting can lead topological superconductivity in these materials \cite{parhizgar-TSC}.

Quantum anomalous Hall (QAH), theoretically predicted in Ref. \cite{science329}, was experimentally realized in magnetically doped TI thin films \cite{science340, prl113137201, nature10731}. In fact, magnetic ordering in TI thin films leads to a band topology at zero external magnetic field. Although this effect has been confirmed by other experimental groups, the theoretical background especially the mechanism emerging such a ferromagnetism is still under debate \cite{science329, peixoto, prb92201304, prb78195207, prb89165202, sciadv1, prl114, natc6, natc7}. So investigation of the effect of impurity on density of states in this material is so important.
Since the QAH experiment has been done at zero chemical potential, the role of impurity bands in both transport properties and also mechanism of coupling between impurities is of much importance when the chemical potential lies inside the gap.

In this work, we show how a single impurity influences the band structure of TI thin film. By using Green's function formalism and making a $T$-matrix expansion \cite{prb73125411, prl102, balatsky, balatsky2, Jonas, Annica, prb94}, we analytically explore the local density of electron states (LDOS) and spin-LDOS near non-magnetic and magnetic impurities, respectively, and show the emergence of new states in the spectrum due to the presence of these impurities. Since these new states can have an important effect on the transport properties, such as the QAH effect, we study their behavior as a function of the system parameters such as type of impurity and its potential 
and also TI's parameters like as the electric potential difference and hybridization between surfaces. In addition, we indicate that one can electrically tune the effect of impurity on the LDOS and spin-LDOS of the system and so manipulate the electronic and spin texture of the system by an applied electric potential.

The impurities in TIs can be located in the bulk as well as on the surface, however, since TIs usually should be in proximity with superconductors or ferro-magnets for possible applications, locating the impurities on the surface provide more applications \cite{Sessi-2014}. Hence, in this work we restrict ourselves to surface impurities. Also, we assume that the impurities are sufficiently dilute such that multiple scatterings between impurities can be neglected and which allows us to restrict the study to the single impurity picture.

The arrangement of this paper is as follows. Firstly, in section \ref{sec2}, the effective model Hamiltonian of 2D TI thin film is presented. We state how the Green's function of the system is calculated using the $T$-matrix approach as well as the calculations of the LDOS and the spin-LDOS in section \ref{sec3}. In section \ref{sec4}, we focus on the obtained results in the two cases (i) non-magnetic and (ii) magnetic impurities, and describe what happens in the LDOS. Moreover, for the sake of straightforwardness, we introduce a useful model named \emph{the two atoms model} to describe some of our results. Finally in section \ref{sec5}, the paper is briefly concluded and summarized. Furthermore, details of calculations are provided in Appendices.

\section{Model Hamiltonian}\label{sec2}
The 2D effective Hamiltonian for the surface states of the TI thin film around the $\Gamma$ point can be written as \cite{nature584,Hamiltonian}

\begin{align}
\label{eq:1}
   H_0(k)=-D\,k^2\sigma_0\otimes\tau_0+[\hbar v_{F} (\sigma\times {\bf k})\cdot \hat z + V\sigma_0]\otimes\tau_z+\Delta\sigma_0\otimes\tau_x,
\end{align}
where $\sigma , \tau$ denote the Pauli matrices in spin and surface space, respectively, $D$ is a coefficient that represents the electron-hole asymmetry in the system, $v_F$ is the Fermi velocity of the surface electrons, ${\bf k}$ denotes the wavevector ($k=|{\bf k}|$) of the surface electrons, and $ V $ is the potential difference between different surfaces. The last term $\Delta$, represents hybridization between surface states, which in general is of the form $\Delta_0-\Delta_1 \,k^2$ (see Fig. \ref{fig:1}(a)). While $\Delta_0$ is a simple tunneling between different surfaces, $\Delta_1$ may result in a quantum phase transition from a QSH insulator to a normal insulator for TI thin films whenever $\Delta_0\Delta_1>0$ and $V<\hbar v_F\sqrt{\Delta_0/\Delta_1}$ \cite{effective,electrically}. In this work, we restrict ourselves to the low energy regime of the Hamiltonian and keep the terms up to linear order in $k$. The electron-hole asymmetry term $D$ usually has a negligible effect on the electronic properties while the effect of $\Delta_1$ may be important for finite size nano-ribbons of TI thin film where changes in the topology can result in the existence of zero-energy edge states \cite{effective,nature438,electrically}.
Table \ref{tab:1} shows the experimental parameters of Bi$_2$Se$_3$ which has been achieved by fitting the parameters of Hamiltonian Eq. \eqref{eq:1} with ARPES data \cite{nature584}.
\begin{center}
\begin{table}[bp]
\caption{Experimental parameters for Bi$_2$Se$_3$ film  \cite{nature584}.}
\label{tab:1}
\begin{tabular}{ |c|c|c|c|  }
 \hline
 No. of QLs& $ v_F $ ($ 10^5 ms^{-1} $)  & $ \Delta_0 $ (meV)& $ \Delta_1 $ (eV\AA$^2$) \\
 \hline
 3   & 4.81   & 69 &   18.0\\
 4&   4.48  & 35   & 10.0\\
 5 & 4.53 & 2.05 &  5.0\\
 \hline
\end{tabular}
\end{table}
\end{center}

In the basis of surface and spin space, the matrix form of the Hamiltonian can be expressed by
\begin{align}
H_0(k)&
\nonumber\\=&
	\begin{bmatrix}
		 V & i\hbar v_Fk e^{-i\phi_{k}} & \Delta & 0 \\
             	-i\hbar v_Fk e^{i\phi_{k}} & V & 0 & \Delta \\
               	\Delta & 0 & -V & -i\hbar v_Fk e^{-i\phi_{k}} \\
               	0 & \Delta & i\hbar v_Fk e^{i\phi_{k}} & -V \\
	\end{bmatrix},
\label{eq:2}
\end{align}
where $\tan\phi_{k}=k_{y}/k_{x}$. The energy dispersion of this Hamiltonian is given by
\begin{align}
\label{eq:3}
E(k)=&
	\pm
	\sqrt{
		\Bigl(
			\hbar \, v_{F} \, k\mp V
		\Bigr)^2
		+
		\Delta^2
	}
	.
\end{align}
This dispersion relation is illustrated in Fig. \ref{fig:1}(b). The structure inversion asymmetry (SIA) term, $V$, between the two surfaces can be resulted from interaction between the TI material and the substrate or by an electric field applied perpendicular to the surface of the thin film \cite{nature584}. The plots in Fig. \ref{fig:1}(b) clearly demonstrate that the SIA generates a band splitting analogous to the Rashba splitting.

\section{Impurity scattering}\label{sec3}
The purpose of this section is to give a detailed account for the calculations of the impurity scattering and its effects on the local electronic structure. As we are interested in scattering of both non-magnetic and magnetic impurities, we outline the general features of the calculations using a generic short ranged scattering potential ${\bf U}(\bfr)={\bf u}_{0/m}\delta(\bfr-\bfr_0)$. Then, we write the total Hamiltonian as
\begin{align}
\label{eq:4}
H=&
	H_0(k)+{\bf U}(\bfr)
	.
\end{align}
Specifically, from now on, we use the notation ${\bf u}_0=u\sigma^0$ and ${\bf u}_m={\bf m}\cdot\boldsymbol{\sigma}$ for the non-magnetic and magnetic impurity, respectively, where $\sigma^0$ is the $2\times2$ identity matrix whereas $\boldsymbol{\sigma}$ is the vector of the Pauli matrices. Moreover, $u$ provides the strength of the non-magnetic scattering potential while ${\bf m}$ represents both the strength of the magnetic scattering potential as well as the direction of the magnetic moment.
In the present set-up, we have assumed that the impurity is located on the upper surface, however, we are interested in its induced effects on both upper and lower surfaces.

The impurity scattering modified electronic structure is addressed by using real space Green's functions (GFs) ${\bf G}(\varepsilon;\bfr,\bfr')$ and we relate the LDOS $\rho(\bfr,\varepsilon)$ and local magnetic texture (spin-LDOS) $\boldsymbol{\rho}_\pm(\bfr,\varepsilon)$ through the relations
\begin{subequations}
\begin{align}
\label{eq:5a}
\rho(\bfr,\varepsilon)=&
	-\frac{1}{\pi}{\rm Im\ Tr\ }{\bf G}(\varepsilon;\bfr,\bfr)
	,
\\
\label{eq:5b}
\boldsymbol{\rho}_\pm(\bfr,\varepsilon)=&
	-\frac{1}{2\pi}{\rm Im\ Tr\ }(\sigma^0\pm\boldsymbol{\sigma}){\bf G}(\varepsilon;\bfr,\bfr)
	.
\end{align}
\end{subequations}
Note that one can assume the magnetic impurity along one spin direction and calculate the spin-LDOS in any other spin direction, which in experiment is equivalent to spin polarization direction of scanning tunneling microscopy (STM) tip. In the above equation, $\boldsymbol{\sigma}$ shows the direction of the spin-LDOS measurement.
\\
We consider the effects of impurity scattering by employing the $T$-matrix approach to calculate the modified, or, dressed GF \cite{prl102,balatsky,balatsky2,Jonas2006,she2013,Jonas,Annica,Jonas2015}. Through a straightforward calculation, we have obtained the dressed GF (henceforth we set $r_0=0$)

\begin{align}
\label{eq:6}
{\bf G}^r(\varepsilon,\bfr,\bfr')=
	{\bf G}_{0}^r(\varepsilon,\bfr,\bfr')
		+
		{\bf G}_{0}^r(\varepsilon,\bfr,0)
		{\bf T}(\varepsilon)
		{\bf G}_{0}^r(\varepsilon,0,\bfr')
		,
\end{align}
where $G_{0}^r$ is the unperturbed (bare) retarded GF for the pristine material whereas the $T$-matrix is defined by
\begin{align}
\label{eq:7}
{\bf T}(\varepsilon)=&
  	{\bf U}
	+
	{\bf U}
	{\bf G}_{0}^r(\varepsilon;0,0)
	{\bf T}(\varepsilon)
	=
	\Bigl(
		{\bf U}^{-1}
		-
		{\bf G}_0^r(\varepsilon,0,0)
	\Bigr)^{-1}
	.
\end{align}
This expression represents the propagation of an excitation in the perfect lattice in which scattering, to arbitrary order, takes place at the single impurity represented by ${\bf U}$ \cite{Gonis}.

We relate the real space GF to the reciprocal space properties through the Fourier transform (${\bf R}={\bf r}-{\bf r'}$)
\begin{align}
\label{eq:8}
{\bf G}_0^r(\varepsilon,{\bf R})=&
 	\frac{1}{\Omega_{BZ}}
	\int d{\bf k}
		 e^{i{\bf k}\cdot{\bf R}}
		 {\bf G}_0({\bf k})
,
\end{align}
where ${\bf G}_0(\varepsilon,{\bf k})=[\varepsilon-H_0({\bf k})]^{-1}$ and $ \Omega_{BZ} $ shows the first Brillouin zone area.

In the following, we consider two different types of impurities separately, namely non-magnetic and magnetic impurity.
First, we consider a local non-magnetic impurity on the upper surface of the thin film for which ${\bf u}_0$ is a $4 \times 4$ matrix in the spin and surface space.

\begin{align}
\label{eq:9}
{\bf u_0}=&
	u
	\delta(r)
	\begin{bmatrix}
	    \sigma_0 &\vdots   & 0 \\
	    \dots & \dots & \dots \\
	    0 &\vdots & 0 \\
	\end{bmatrix}
	.
\end{align}
Hence, for the non-magnetic impurity we obtain the $T$-matrix
\begin{align}
\label{eq:10}
{\bf T}_0=&
	\frac{u}{1-u g_{11}}
	\begin{bmatrix}
	    \sigma_0 &\vdots   & 0 \\
	    \dots & \dots & \dots \\
	    0 &\vdots & 0 \\
	\end{bmatrix}
	,
\end{align}
where
\begin{align}
\label{eq:11}
g_{11}=&
	-\frac{2\pi}{\Omega_{BZ}}
	\sum_{s=\pm}
	\int^{k_{c}}_0dk k
		\frac{a_s(\gamma+isV)}{\hbar^2v_F^2k^2-(V-is\gamma)^2}
	,
\end{align}
is the upper left component of ${\bf G}_0(\varepsilon,0,0)$ (see Eq. \eqref{eq:A3}). Here, $k_c$ is the cut-off wave-vector, $\gamma^2=\Delta^2-\varepsilon^2$ and $a_\pm=(\varepsilon/\gamma\pm i)/2$. Then, the corresponding LDOS is given by
\begin{align}
\label{eq:12}
\rho=&\frac{-1}{\pi}{\rm Im}
	\biggl[
	2g_{11}
	-
	8\pi^2\alpha^2u
	\frac{F_0-F_1}{1-u g_{11}}
	\biggr],
\end{align}
where $\alpha=1/\hbar^2 v_F^2 \Omega_{BZ}$, $F_0=[\sum_{s=\pm} s \; a_{-s} (V+is \gamma ) K_{0}^{s}]^2$ and $F_1=[\sum_{s=\pm}sa_{-s}K_{1}^{s}/\sqrt{-1/(V+is \gamma )^2}]^2$, whereas
\begin{align}
\label{eq:13}
K_{0}^s=
	K_0(-ir/x_s),
	&&
K_{1}^s=
	K_1(-ir/x_s),
\end{align}
in which $K_n(x)$ is the modified Bessel function and $x_s=\sqrt{\hbar ^2 v_F^2/(V\pm i \gamma )^2}$, $s=\pm1$.

In the case of a single magnetic impurity located on the upper surface of the thin film, the  scattering potential is given by
\begin{align}
\label{eq:14}
{\bf u}_m=&
	\delta(r)
	\begin{bmatrix}
	    {\bf m}\cdot\boldsymbol{\sigma} &\vdots   & 0 \\
	    \dots & \dots & \dots \\
	    0 &\vdots & 0 \\
	\end{bmatrix}
	.
\end{align}
For a magnetic moment polarized in the $\hat{\bf z}$-direction,  ${\bf m}=m_z\hat{\bf z}$, the $T$-matrix acquires the form
\begin{align}
\label{eq:15}
{\bf T}_m=&
	\begin{bmatrix}
		\frac{m_z}{1-m_zg_{11}} & 0 & 0 & 0  \\
		0 & \frac{-m_z}{1+m_zg_{11}} & 0 & 0 \\
		0 & 0 & 0 & 0 \\
		0 & 0 & 0 & 0 \\
	\end{bmatrix}.
\end{align}
Hence, the corresponding spin-LDOS in the upper surface for the up and down spin is given by
\begin{subequations}
\label{eq:16}
\begin{align}
\label{eq:16a}
\rho_{\uparrow}^{z,u}=&\frac{-1}{\pi}{\rm Im}
	\biggl[
	g_{11}
	-
	4m^2\pi^2\alpha^2
	\bigg(	
		\frac{F_0}{1-m g_{11}}
		-
		\frac{F_1}{1+m g_{11}}
	\biggr)
	\biggr]
	,
\\
\label{eq:16b}
\rho_{\downarrow}^{z,u}=&\frac{-1}{\pi}{\rm Im}
	\biggl[
	g_{11}
	+
	4m^2\pi^2\alpha^2
	\biggl(
		\frac{F_0}{1+m g_{11}}
		-
		\frac{F_1}{1-m g_{11}}
	\biggr)
	\biggr]
	.
\end{align}
\end{subequations}
Analogous expressions are also obtained for the lower surface (see Appendix \ref{aapp-zpol}). The same calculations can be simply done for other impurity's alignments and other spin directions measurement.

\section{Results}\label{sec4}
In this section, we present our results for the impurity scattering's influences on the surfaces of TI thin film. We have not restricted our study to a single thickness and compared the results for $4$ and $5$ QLs of Bi$_2$Se$_3$. We study the impurity scattering effects with respect to the potential drop $V$ between the surfaces. In all our plots, we have calculated the LDOS and spin-LDOS at a distance of 30 nm away from the scattering center.

\subsection{Non-magnetic impurity}

\begin{figure}[t]
\includegraphics[width=.99\columnwidth]{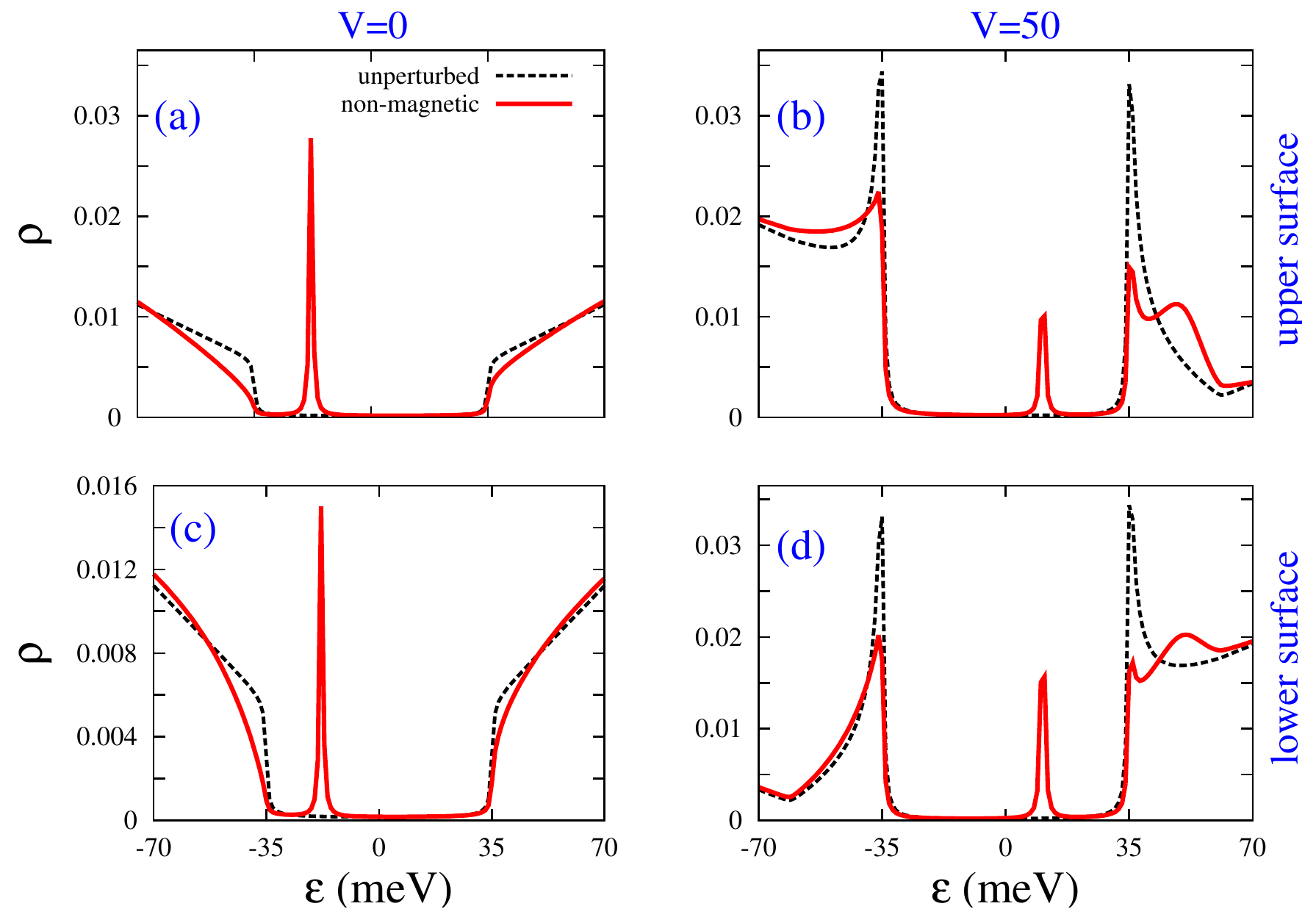}
\caption{(Color online) The LDOS of the upper (a, b), and lower (c, d) surface for $\Delta=35$ meV, $u=100$ eV, $r=30$ nm and different values of voltage $V=0$ (a, c), $50$ meV (b, d). The LDOS of the unperturbed surface (dashed) is included for reference.}
\label{fig:2}
\end{figure}

We begin our survey by studying the effects arising from the non-magnetic impurity. In Fig. \ref{fig:2}, we plot the LDOS for the upper (a, b) and the lower (c, d) surfaces, for two different values of asymmetry potential $V=0$ (a, c) and $V=50$ meV (b, d). We have also included the LDOS for the unperturbed surfaces (dashed line) for reference. As expected, the hybridization between the surfaces creates a density gap whenever $|\varepsilon|<\Delta$ in the unperturbed LDOS. The density of states for the bare system related to Eq. \eqref{eq:3} is given by $D(\varepsilon)=\sum_{i}\varepsilon k_{i}\Theta(\varepsilon^2-\Delta^2)/(\hbar v_Fk_{cr}\pm V)$ where $k_{i}$ refers to $k$ points in which $\varepsilon=E(k)$ is given by $\hbar v_F k_{i}=\sqrt{\varepsilon^2-\Delta^2}\pm V$. As it is obvious, the Van-Hove singularities occur at $\hbar v_F k=\pm V$ for nonzero $ V $ and there is no Van-Hove singularity at $ V=0 $.
In the case with vanishing asymmetry potential, $V=0$, the LDOS grows linearly with energy outside the energy gap and we notice that the electron-hole symmetry is preserved. However, under a finite asymmetry potential, the Van-Hove singularities emerge at the band edges and also the electron-hole symmetry breaks within the individual surface. Nevertheless, since the energy band dispersion is a combination of the properties from both surfaces, the overall electron-hole symmetry is preserved also for finite $V$. The LDOS in the lower surface is obtained by changing $ \varepsilon \rightarrow -\varepsilon $ in the LDOS of the upper surface.

In the presence of a single non-magnetic impurity, a single state emerges in the gap, see the in-gap peaks in the red solid lines in Fig. \ref{fig:2}. This new peak is related to the poles of Eq. \eqref{eq:12} (at $1-u g_{11}=0$) such that the position and height of this peak depend on the parameters of the system, e.g., $u,\, \Delta,\, V$. By increasing the applied voltage, at low voltages, the peak shifts to higher energies. Therefore, there exits a controllable way to
behave TI thin films as a semiconductor doped by acceptors
or donors\cite{phosphorene1, phosphorene2, phosphorene3}.

This shows that, in such system, one is able to control and tune electrically the scattering effects of impurity. Furthermore, Van-Hove singularities will be softened in the presence of non-magnetic impurity and, in addition, the states outside the gap which has been linear in terms of energy would change.

Also, one can see at $V=0$ shown in Fig. \ref{fig:2} (a,c) that the upper surface has a stronger in-gap impurity peak in comparison with the lower surface, however, this order of the their relative strengths may not hold for finite $V$. This suggests that, for a finite value of the asymmetry potential, one may see a stronger effect of impurity in the lower surface, although the impurity is located on the upper surface.

\subsection{Magnetic impurity}
Now we turn our attention to the effect of a magnetic impurity on the spin-LDOS. The impurity's moment is aligned in $\alpha(=\hat{x},\hat{y},\hat{z})$-direction and the spin-LDOS can be measured in an arbitrary $\beta(=\hat{x},\hat{y},\hat{z})$-direction of spin which has been denoted in Eq. \eqref{eq:5b} by  $\boldsymbol{\sigma}$. Also in general, this measurement depends on the distance vector ${\bf r}$, shown the position of measurement from the located impurity at $r_0=0$. In this work, we have fixed the magnitude of the distance $r=30$ nm and we have considered two different directions for ${\bf r}$ namely $ \bf{\hat{x}} $ and $ \bf{\hat{y}} $. For simplicity, we exhibit such situation of under study by $ {\bf{\cal{F}}}_{\bf{\alpha},\bf{\beta}}^{\bf{\hat{x}}/\bf{\hat{y}}} $.

\begin{figure}[t]
\centering
\includegraphics[width=.95\linewidth]{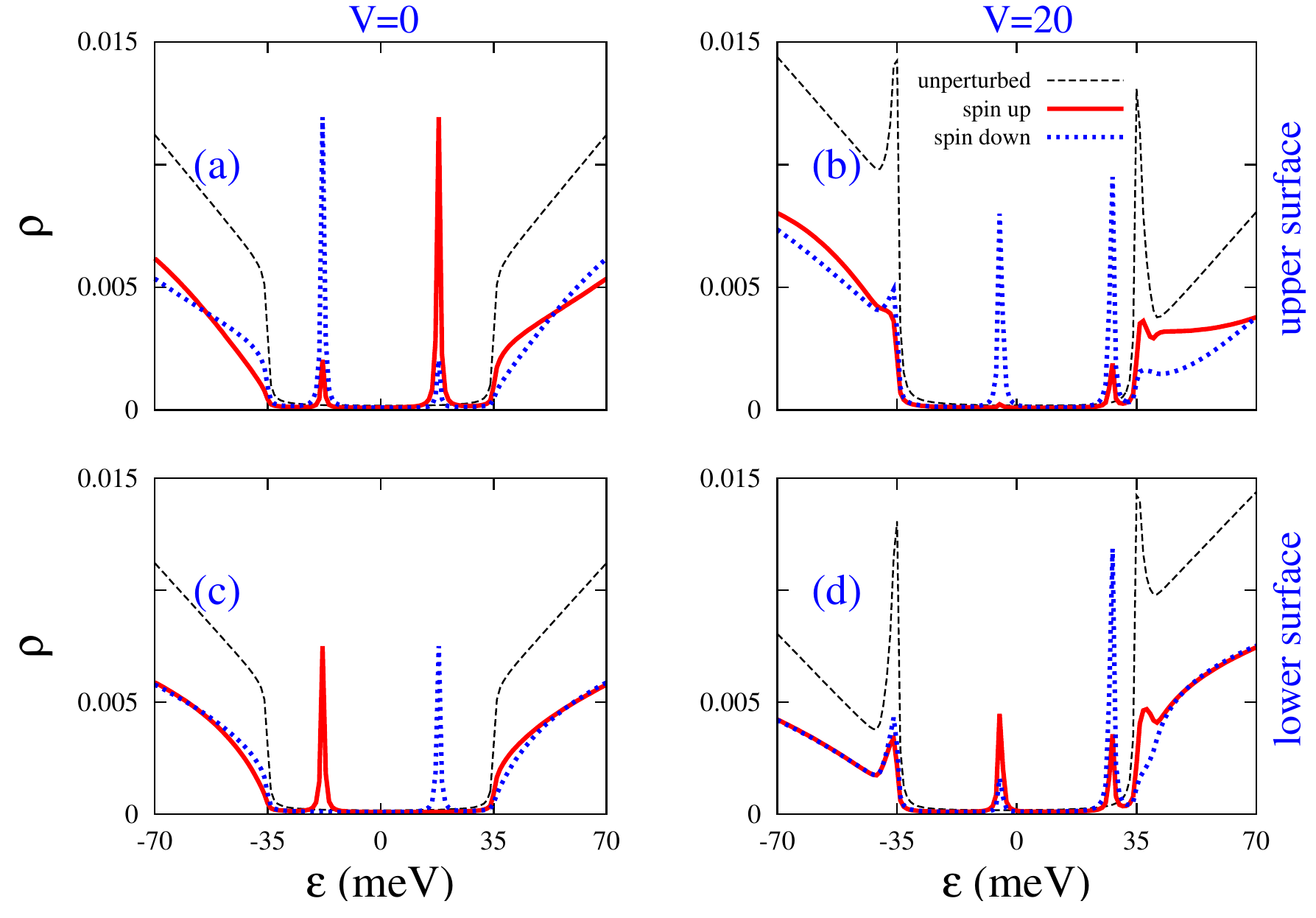}
\caption{\small{(Color online)The effect of $ \bf{\hat{z}} $-polarized magnetic impurity located on the upper surface of TI thin film on the spin-LDOS of upper/lower surface, for $\Delta=35$ meV, $ m=100$ eV, $r=30$ nm and different values of voltage $ V=0, \; 20 $ meV. Dashed line refers to the LDOS of unperturbed system. Red solid line and dot blue line show spin up and down LDOS respectively.}}
\label{fig:3}
\end{figure}

Let us first focus on the cases of a $\bf{\hat{z}} $-polarized magnetic impurity (see Eq. \eqref{eq:16}) and also assume the measurement to be done in $\hat{z}$ spin direction. The measurement of spin-LDOS in this situation is independent of the ${\bf r}$ direction and so it can be labeled by  $ {\bf{\cal{F}}_{z,z}} $. Fig. \ref{fig:3} shows the spin-LDOS of both upper and lower surfaces for a 4QLs Bi$_2$Se$_3$ TI thin film in situation $ {\bf{\cal{F}}_{z,z}} $. The gap parameter according to Table \ref{tab:1} is $35$ meV and two different asymmetry potentials $V=0, \; 20$ meV have been compared.
As the magnetic impurity breaks the degeneracy of spin states, two peaks appear inside the gap region. Roughly speaking, one can say each of these states should belong to one spin, however, since TI thin film has a strong Rashba spin-orbit coupling which couples different spins, we would have four peaks at two energies.

For $V=0$ (Fig. \ref{fig:3}(a, c)), the four peaks appearing in these figures have the symmetry that any spin-LDOS, including the peaks in the positive energies, is equivalent to the opposite spin in the negative energy. For the lower surface at $V=0$, shown by Fig. \ref{fig:3}(c), just two peaks emerge according to different spins and there would be no state for opposite spin there. This is the same for other thicknesses at $V=0$. As one increases the potential $V$, both new states inside the gap would shift to higher energies (Fig. \ref{fig:3}(b, d)). However, as it will be shown in Fig. \ref{fig:em}, at higher voltages impurity states emerge at lower energies. For the lower surface, Fig. \ref{fig:3}(d), the four peaks come back and they occur exactly on the same energies as the upper surface. Also by increasing $V$ the symmetry between different spins and energies breaks for both surfaces.

\subsubsection*{Analogy with two atoms model:}
To describe our results, we want to make an analogy of our system with a molecule model containing of two single atoms \cite{cheraghchi} ($A$ and $B$ with one orbital per site) connecting to each other by the hopping energy $t$ . Let us consider on-site energies to be as $E_A=-M$ and $E_B=M$.

The molecular energy eigenvalues of the system would be at $ E_\pm=\pm \, \sqrt{t^2+M^2} $ while the corresponding eigenstates are obtained for bonding state as $ \psi_+=\beta \phi_A+\alpha \phi_B$ and for anti-bonding state as $\psi_-=\alpha \phi_A-\beta \phi_B$ where $ \alpha=\sin\phi$ and $\beta=\cos\phi$ and $\tan2\phi=t/M $. An asymmetry of the on-site energies results in an asymmetry in the LDOS on bonding and anti-bonding states $E_\pm$,

\begin{subequations}
\label{eq:18}
\begin{align}
LDOS(E,A)=&
	\Bigl(
		\alpha^2 \, \delta(E-E_+)+\beta^2 \, \delta(E-E_-)
	\Bigr)
	,
\\
LDOS(E,B)=&
	\Bigl(
		\alpha^2 \, \delta(E-E_-)+\beta^2 \, \delta(E-E_+)
	\Bigr)
	.
\end{align}
\end{subequations}
In the limit of large energy difference (ionic limit), $ \mid t \mid \ll M $, coefficients are nearly approximated by $\alpha \approx t/2M$ and $\beta \approx 1-0.5 (t/2M)^2$, which show that the peaks in the LDOS are more localized on the energies close to the on-site energies of the given atom.

\begin{figure}[t]
\includegraphics[width=1.0\linewidth]{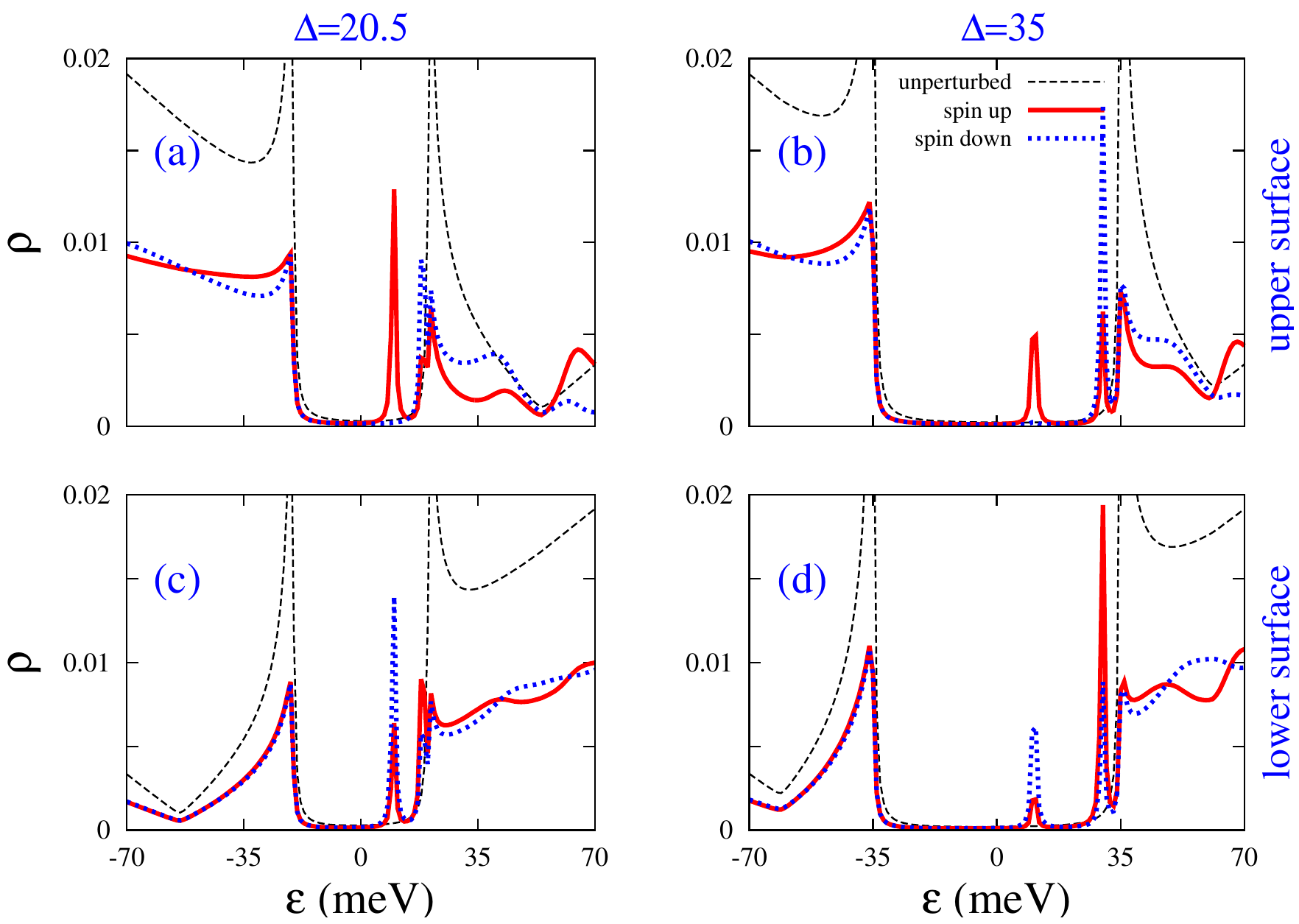}
\caption{\small{(Color online)The effect of $ \bf{\hat{z}} $-polarized magnetic impurity located on the upper surface of TI thin film on the spin-LDOS of upper/lower surface, for $\Delta=20.5, 35$ meV, $m=100$ eV $r=30$ nm $V=50$ meV. Dashed line refers to the LDOS of unperturbed system. Red solid line and dot blue line show spin up and down LDOS respectively.}}
\label{fig:4}
\end{figure}

In terms of the analogy of A and B sites, representing the spin up and down in our model where the magnetic impurity generate different energies of different spins which plays the role of $M$ in two-atoms model while the Rashba spin-orbit coupling is represented by $t$. The two-atoms model shows why there are four peaks related to the spin up and down at the bonding and anti-bonding energies. At $V=0$ (Fig. \ref{fig:3}
(a, c)), there is no on-site energy on either surfaces, hence, the impurity states inside the gap can act
like as an individual atom which leads to the new peaks inside the gap which are symmetrically distributed around the Fermi level at the energies $\pm E$. Also, consistent with our two atoms model, the LDOS for the spin-up (down) peak at the bonding (anti-bonding) energy equals the LDOS of opposite spin at the opposite energy. An increase in magnetic moment of impurity leads to an enhanced spin-LDOS on the bonding (anti-bonding) energy for spin up (down). Furthermore, in comparison with the upper surface, the up and down spin-polarized impurity states inside the gap are interchanged with each other in the lower surface as an effect of the negative Rashba coupling in the Hamiltonian at this surface (Fig. \ref{fig:3} c). As we increase the bias potential $V$, the on-site energies attributed to the upper and lower surface are asymmetric which may cause an asymmetry in LDOS and a
shift in the bonding and anti-bonding energies. Based on the two atoms model and as long as the impurity states are localized in the gap, the down-spin LDOS on the anti-bonding state is enhanced in the upper surface while the up-spin LDOS
on the bonding state is enhanced in the lower surface. By further increasing $V$, the impurity states move to each other
and mixed with states of the bare system electrons outside the gap where the two atoms model is no longer valid.

To investigate the effect of $\Delta$ on the spin-LDOS, we depicted the spin-LDOS for 3 and 4 QLs Bi$_2$Se$_3$ TI thin films at $V=50$ meV in Fig. \ref{fig:4}. As the previous figure, we assumed the impurity to be again $\hat{z}$-polarized and calculate the spin-LDOS in the same spin direction, i.e. $ {\bf{\cal{F}}_{z,z}} $ situation. First, by comparing Fig. \ref{fig:4}(b, d) with Fig. \ref{fig:3} in which all of them are for 4QLs Bi$_2$Se$_3$, it is observed that by increasing $V$ to higher values than $\Delta$ in Fig. \ref{fig:4}, the peaks shift to the band edges giving rise hybridization of the magnetic localized states with the TI states. This hybridization causes to occur some distortions in the conduction band.

Also a comparison between 3QLs and 4QLs in this figure shows that by reducing $\Delta$, the energy position of the two appeared peaks becomes closer to each other \cite{balatsky}.
Besides, it is worth to mention that the position and LDOS value of these new states inside the gap depend on the value of $m$. For very low magnetic strength, no peak would pop up inside the gap, as one increases this value, the peaks would appear and their LDOS value increases by $m$ and they become close to each other. In the magnetic impurity case, it should be noted that we have not considered any scattering of electrostatic potential (${\bf u}_0$), so at $m\rightarrow 0$, there is no impurity state. In a real situation which magnetic impurity could generate a non-magnetic potential as well, one should consider the general form of potential ${\bf u}_0+{\bf u}_m$\cite{Annica}.

\begin{figure}[t]
\centering
\includegraphics[width=1\linewidth]{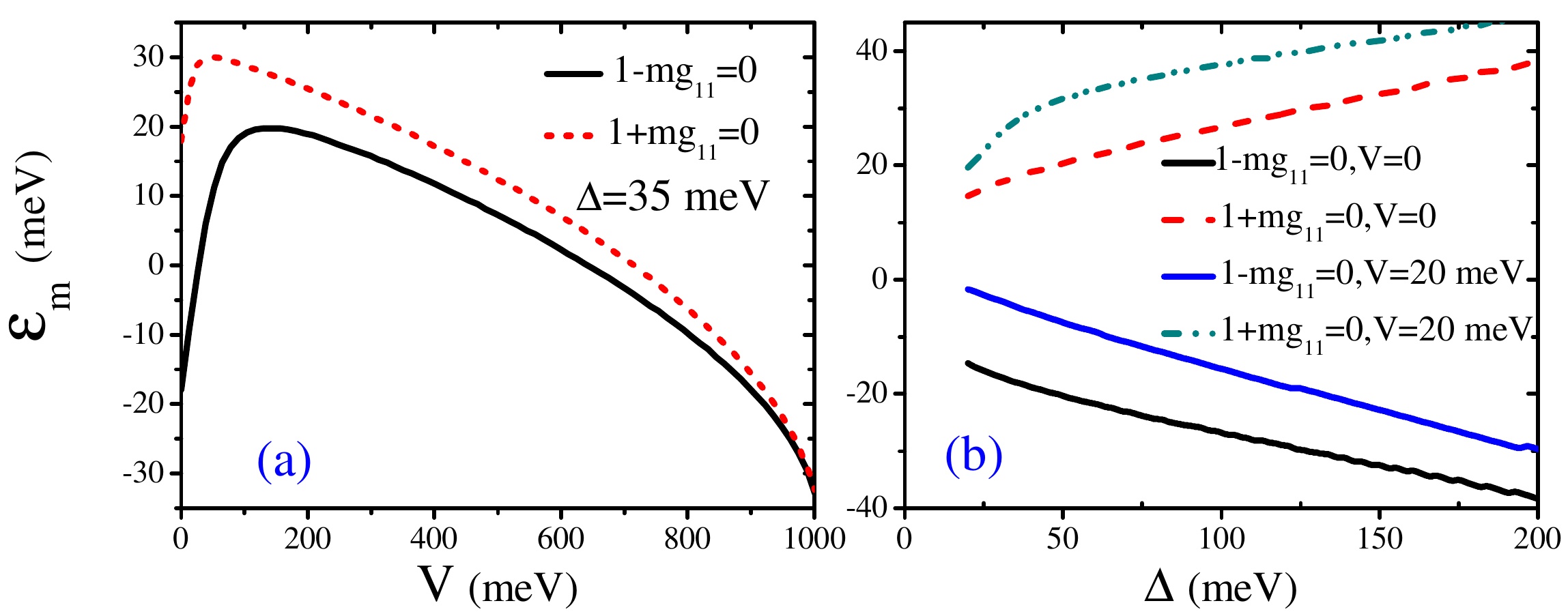}
\caption{\small{(Color online) The energy of bound state induced by single magnetic impurity inside the gap, with respect to (a) biased potential $V$ (b) gap size $\Delta$. }}
\label{fig:em}
\end{figure}

Figure \ref{fig:em} shows the bound state energies of magnetic impurity ($\varepsilon_{m}$) with respect to the biased voltage $V$ and size of the gap $\Delta$. These energies are related to the zeros of denominator of the $T$-matrix in Eq. \eqref{eq:16a},\eqref{eq:16b} in which $1\pm mg_{11}=0$. The behavior of energy position according to the non-magnetic impurity is equivalent to the $\varepsilon_{m-}$ in which $1-mg_{11}=0$. In the panel (a) where we depict $\varepsilon_m$ with respect to $V$, two peaks start from symmetric energies at $V=0$ (\textit{i.e.} $|\varepsilon_{m-}|=|\varepsilon_{m+}|$) and they increase with respect to $V$. More increasing of the biased voltage causes to decrease both of these peak's energies and reach to minus energies and would touch the valence band edge at a critical voltage (here $\sim 1$eV). After this critical voltage there would be no impurity bound state inside the gap. Fig. \ref{fig:em} (b) depicts the bound state's energies with respect to the size of the gap from $20$ meV up to $200$ meV and for two different voltages $0, 20$ meV. As one can see, at $V=0$, the peaks occur at the symmetric energies against the point $\varepsilon=0$, however, both energies shift to higher values for $V=20$ meV.

\begin{figure}[t]
\centering
\includegraphics[width=1\linewidth]{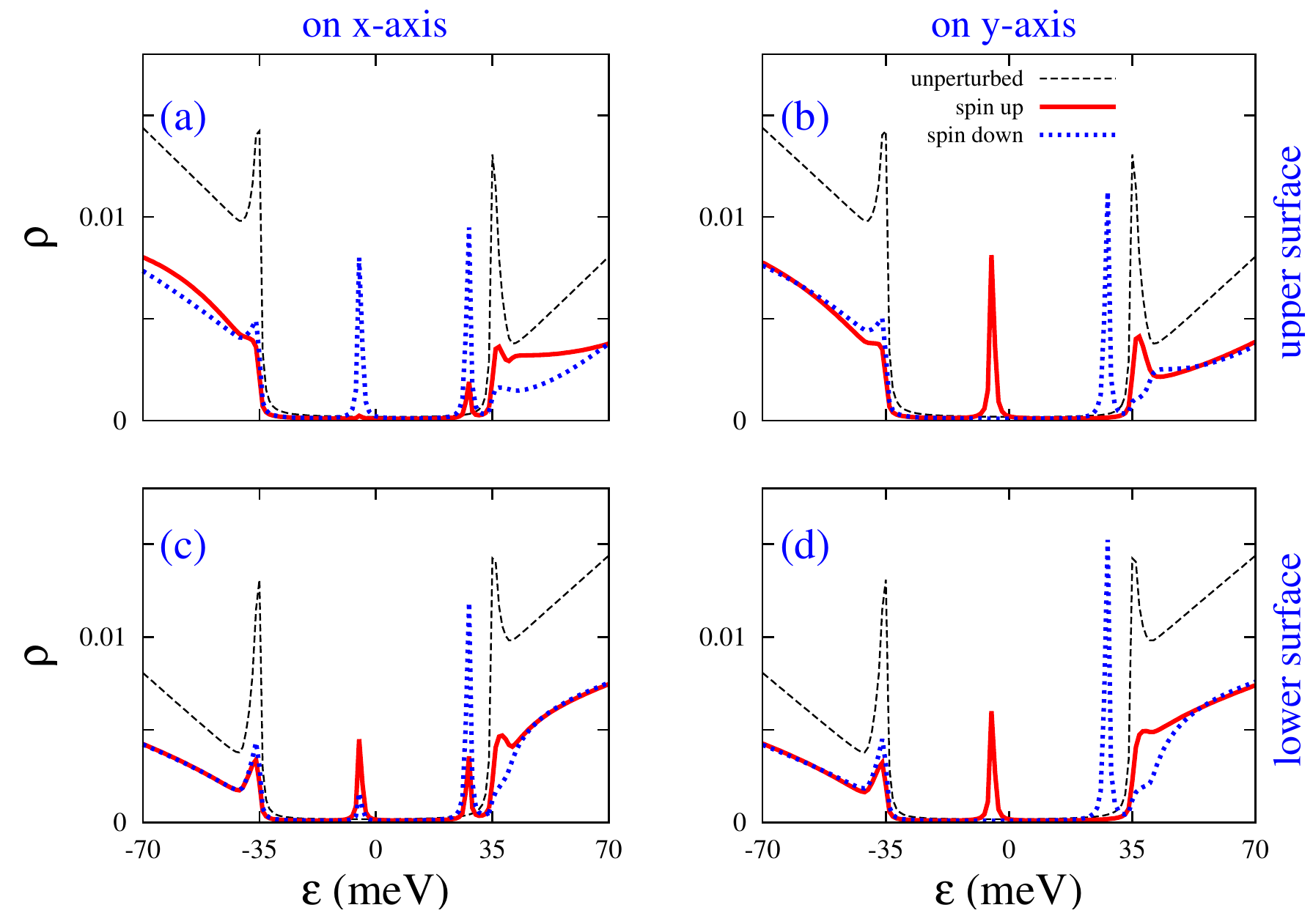}
\caption{\small{(Color online)The effect of $ \bf{\hat{x}} $-polarized magnetic impurity located on the upper surface of TI thin film on the spin-LDOS of the upper/lower surface on $\hat{x}/\hat{y}$ axis. $\Delta=35$ meV, $V=20$ meV, $m=100$ eV and $r=30$ nm. Dashed line refers to the LDOS of unperturbed system. Red solid line and dot blue line show spin up and down LDOS respectively.}}
\label{fig:5}
\end{figure}

Furthermore, we consider the $ \bf{\hat{x}} $-polarized magnetic impurity to be located on the upper surface and calculate the spin-LDOS in the same spin direction. In this case, the results would be spatially anisotropic and so we have presented our result for two different spatial directions $ \hat{x} $ and $ \hat{y} $, which such situations are shown by $ {\bf{\cal{F}}_{x,x}^{ \bf \hat{x}/\hat{y}}} $(see Appendix \ref{aapp-xpol}). We present the effect of $\bf{\hat{x}}$-polarized magnetic impurity in Fig. \ref{fig:5} which shows the spin-LDOS of the thin film for both surfaces. Here, we chose the tunnelling, $\Delta=35$ meV, and asymmetry potential, $V=50$ meV. The anisotropy of spin-LDOS for different spatial directions is visible in this figure. Comparing Fig. \ref{fig:3}(b, d) with Fig. \ref{fig:5}(a, c) clarifies that the spin-LDOS due to the situations $ {\bf{\cal{F}}_{z,z}} $ and $ {\bf{\cal{F}}_{x,x}^{ \bf \hat{x}}} $ gives us the same results.
In addition, our calculations show equality of the spin-LDOS results of two situations $ {\bf{\cal{F}}_{x,x}^{ \bf \hat{x}}} $ and $ {\bf{\cal{F}}_{y,y}^{ \bf \hat{ y}}} $. Also two other situations $ {\bf{\cal{F}}_{x,x}^{ \bf \hat{y}}}$, ${\bf{\cal{F}}_{y,y}^{ \bf \hat{x}}} $ result in the same spin-LDOS.
This can be described by the symmetry of the Hamiltonian where the Rashba term couples the electron's momentum to their spin and so any rotation applied to both spin and spatial directions won't change any physical quantity.

\begin{figure}[t]
\centering
\includegraphics[width=1\linewidth]{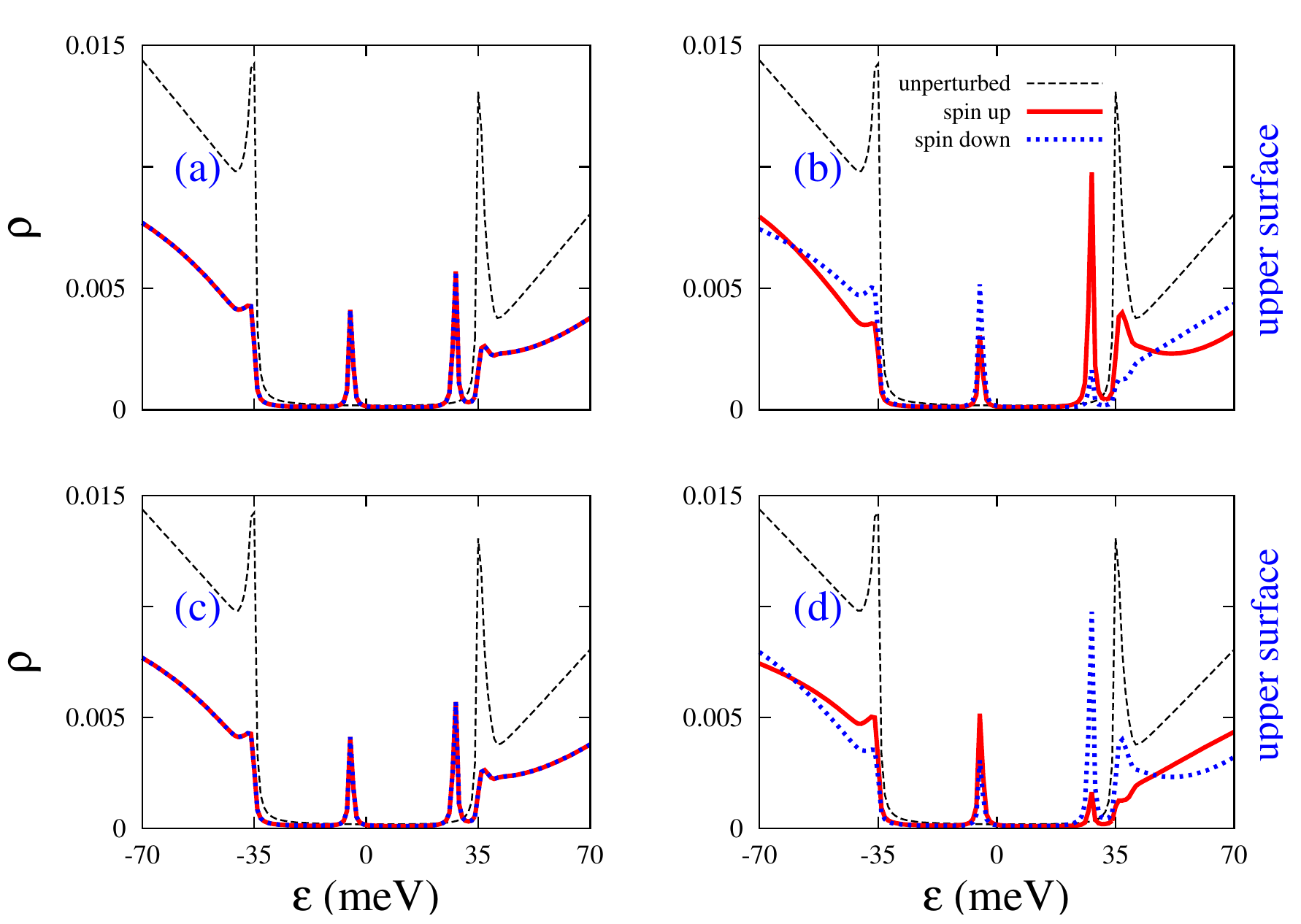}
\caption{\small{(Color online)The effect of different direction-polarized magnetic impurity located on the upper surface of TI thin film on the various directional spin-LDOS of upper surface on $\hat{x}$ axis. Here, we chose $\Delta=35$ meV, $V=20$ meV, $m=100$ eV and $r=30$ nm. Dashed line refers to the LDOS of the unperturbed system. Red solid line and doted blue line show spin up and down LDOS respectively.}}
\label{fig:6}
\end{figure}

\begin{table}[b]
\caption{All the possible situations for magnetic impurity and calculated spin-LDOS}
\label{tab.2}
\begin{tabular}{ |m{3.5cm}|m{4.5cm}|  }

 \hline
 \textbf{
 Cases under study
} & \textbf{
 Related figures
} \\[6pt]
 \hline
   $ \bf{\cal{F}}_{z,z} $, $ \bf{\cal{F}}_{x,x}^{\hat{x}} $, $ \bf{\cal{F}}_{y,y}^{\hat{y}} $ & Fig.3(b) , Fig.5(a) \\[6pt]
  \hline
  $ \bf{\cal{F}}_{x,x}^{\hat{y}} $, $ \bf{\cal{F}}_{y,y}^{\hat{x}} $ & Fig.5(b) \\[6pt]
  \hline
  $ \bf{\cal{F}}_{x,y}^{\hat{x}} $, $ \bf{\cal{F}}_{x,y}^{\hat{y}} $, $ \bf{\cal{F}}_{y,x}^{\hat{x}} $, $ \bf{\cal{F}}_{y,x}^{\hat{y}} $, $ \bf{\cal{F}}_{y,z}^{\hat{x}} $, $\bf{\cal{F}}_{x,z}^{\hat{y}} $, $ \bf{\cal{F}}_{z,y} $ & Fig.6(a), Fig.6(c)\\[6pt]
  \hline
   $ \bf{\cal{F}}_{x,z}^{\hat{x}} $, $ \bf{\cal{F}}_{y,z}^{\hat{y}} $ & Fig.6(b) \\[6pt]
  \hline
  $ \bf{\cal{F}}_{z,x} $ & Fig.6(d)\\[6pt]
 \hline
\end{tabular}
\end{table}

Figure \ref{fig:6} shows some of our findings when the direction of the polarization of the magnetic impurity is not the same as the spin-direction of the spin-LDOS. Here, $\Delta=35$ meV, $V=20$ meV, $m=100$ eV and $ r=30 $nm.  Fig. \ref{fig:6}(a) belongs to the  $ {\bf{\cal{F}}_{x,y}^{ \bf \hat{x}}} $ situation. Besides, Fig. \ref{fig:6}(c) is obtained for the spin-LDOS related to the $ {\bf{\cal{F}}_{y,x}^{ \bf \hat{x}}} $. The plots in these two figures suggest that their behavior in two situations are the same and that both situations leads to equal spin up and down LDOS. This spin-unpolarized result would lead to zero spin susceptibility and equivalently RKKY interaction \cite{prl106} in the related directions since this interaction comes from the spin up/down imbalance of itinerant electrons caused by existence of magnetic impurity \cite{mahroo}.
In addition, we calculate the spin-LDOS related to the $ {\bf{\cal{F}}_{y,z}^{ \bf \hat{x}}} $ and $ {\bf{\cal{F}}_{z,y}} $ situations (where in the later case, symmetry around z-axis causes $ {\bf{\cal{F}}_{z,\beta}} $ to be independent of spatial directions) and no difference between the results of these four situations have been seen.
Fig. \ref{fig:6}(b, d) depict the behavior of the spin-LDOS related to two cases, $ {\bf{\cal{F}}_{x,z}^{ \bf \hat{x}}} $ and $ \bf{\cal{F}}_{z,x} $ respectively. Obviously, comparing them demonstrates that their spin-LDOS would be similar provided that spin up(down) changes to spin down(up). Above all, calculations prove that $ {\bf{\cal{F}}_{y,z}^{ \bf \hat{y}}} $ is achieved by application of $ \pi/2 $-rotation on $\bf{\cal{F}}_{x,z}^{ \bf \hat{x}}$ around z-axis. Briefly, Table \ref{tab.2} shows all the possible situations for fixed $ V $ and $ \Delta $ where we have categorized all possible situations in five different groups.

\section{Conclusion}\label{sec5}
In conclusion, we have investigated the effect of single non-magnetic and magnetic impurity respectively on the LDOS and spin-LDOS of TI thin films. We found analytic results for the Green's function in the real space so one can extract the DOS of the system with favorable experimental parameters. We find that a sufficiently strong potential associated with the single impurity generates states inside the gap. Hence, for many impurities with different potentials one can expect the gap to be filled or, at least, strongly modified. Since interesting experiments such as quantum anomalous Hall effect has been done at zero chemical potential, existence of these new states can have important effect on coupling of magnetic impurities (in QAH experiment) as well as transport properties.

The existence of these new states becomes more important when one considers their relaxation time. The relaxation time of the impurity states is proportional to the inverse of their self-energy which in the first Born approximation is proportional to the bare density of states of the system. The appearance of these new peaks inside the gap indicates that they are stable with relatively long life-times compared to bound states outside the gap of materials known as virtual bound states.

Furthermore, we discuss the symmetries of these new states and categorized them with respect to spin direction of magnetic impurity and spin direction in which spin-LDOS is calculated. In addition, since the band dispersion of TI thin film would be affected by application of an electric field perpendicular to the surface of TI thin film, we showed how one can tune the effect of both magnetic and non-magnetic impurities using this voltage.

\section*{Acknowledgment }
F.P. thanks Amir Sabzalipour and Mahdi Mashkoori for useful discussions and M.Sh. acknowledges Institute for Research in Fundamental Sciences for their hospitality while the last parts of this paper were preparing. J.F. acknowledges support from Vetenskapsr\aa det. H.C. acknowledges the International Center for Theoretical Physics (ICTP) for their hospitality and support in which initial stages of this work was started.
\begin{appendix}

\section{Details of Green's function}
\label{app-GF}
In this section, details of calculation are given more explicitly.
By the Fourier transformation, the unperturbed retarded GF in real space for the TI thin film will be achieved

\begin{align}
\label{eq:A1}
{\bf G}_0^r(\varepsilon,r)=&
	\begin{bmatrix}
	    G_{11} & -G_{21}  & \vdots & G_{13} & -G_{23} \\
	    G_{21} & G_{11} & \vdots & G_{23} & G_{13} \\
	    \dots & \dots & \dots & \dots & \dots\\
	    G_{13} & -G_{23} & \vdots & G_{33} & -G_{43} \\
	    G_{23} & G_{13} & \vdots & G_{43}& G_{33} \\
	\end{bmatrix},
\end{align}
where
\begin{subequations}
\begin{align}
\label{eq:A2}
G_{11}(\varepsilon,r)=&
	-2\pi \alpha \ \sum_{s=\pm} a_{-s}(\gamma-isV)K_{0}^s
	,
\\
G_{21}(\varepsilon,r)=&
	-2\pi i \alpha \  \sum_{s=\pm} \frac{a_{-s}}{\sqrt{\frac{-1}{(V-is\gamma)^2}}}K_{1}^s
	,
\\
G_{13}(\varepsilon,r)=&
	\pi i \alpha \ \frac{\Delta}{\gamma}\ \sum_{s=\pm} s (V+is\gamma)K_{0}^s
	,
\\
G_{23}(\varepsilon,r)=&
	-\pi i \alpha \ \frac{\Delta}{\gamma}\ \sum_{s=\pm} \frac{s}{\sqrt{\frac{-1}{(V+is \gamma)^2}}} K_{1}^s
	,
\\
G_{33}(\varepsilon,r)=&
	-2\pi \alpha \ \sum_{s=\pm} a_s(\gamma-isV)K_0^s
	,
\\
G_{43}(\varepsilon,r)=&
	-2\pi i \alpha \ \sum_{s=\pm} \frac{s a_s}{\sqrt{\frac{-1}{(V+is\gamma)^2}}}K_1^s
	,
\end{align}
\end{subequations}
where $K_{0/1}^\pm $, $\gamma$, $ \alpha $, $a_\pm$ are defined in the main text.

Also the unperturbed on-site GF, $ G_0^r(\varepsilon,0,0)=\bra{0}G_{0}^r(\varepsilon)\ket{0} $, can be obtained by the same Fourier transformation, but this time we need to apply cut-off on k in Eq. \eqref{eq:11}  \cite{modeling,electronic}.

\begin{align}
\label{eq:A3}
{\bf G}_0^r(\varepsilon,0,0)=&
	\begin{bmatrix}
		{g}_{11} & 0 & \vdots & {g}_{13} & 0 \\
		0 & {g}_{11} & \vdots & 0 & {g}_{13} \\
		\dots & \dots & \dots & \dots & \dots \\
		{g}_{13} & 0 & \vdots & {g}_{33} & 0 \\
		0 & {g}_{13} & \vdots & 0 & {g}_{33} \\
	\end{bmatrix},
\end{align}

where the $g_{ij}$:s are defined as
\begin{subequations}
\begin{align}
\label{eq:A4}
g_{11}=&
	-\frac{2\pi}{\Omega_{BZ}}\sum_{s=\pm}\int^{k_{c}}_0 dk \; k\; \frac{a_s(\gamma+isV)}{\hbar^2v_F^2k^2-(V-is\gamma)^2}
	,
\\
g_{13}=&
	i\frac{\pi\Delta}{\gamma \; \Omega_{BZ}}\sum_{s=\pm}\int^{k_{c}}_0 dk \; k\; \frac{s \; (is\gamma+V)}{\hbar^2v_F^2k^2-(V+is\gamma)^2}
	,
\\
g_{33}=&
	-\frac{2\pi}{\Omega_{BZ}}\sum_{s=\pm}\int^{k_{c}}_0 dk \; k\; \frac{a_{-s}(\gamma+isV)}{\hbar^2v_F^2k^2+(isV+\gamma)^2}
	.
\end{align}
\end{subequations}

\section{The spin-LDOS}
\label{app-LDOS}
\begin{widetext}

\subsection{$\bf{\hat{z}}$-polarised magnetic impurity}
\label{aapp-zpol}

In presence of $\bf{\hat{z}}$-polarized magnetic impurity on upper surface, the spin-LDOS relations for lower surface would be
\begin{subequations}
\begin{align}
\label{eq:B1}
\rho_{\uparrow}^{z,l}=&
	g_{33}
	+
		\frac{\pi^2\alpha^2\Delta^2m/\gamma^2}{1-m g_{11}}
		\left[\sum_{s=\pm}(\gamma -i s V) K_0^s\right]^2
	+
		\frac{\pi^2\alpha^2\Delta^2m/\gamma^2}{1+m g_{11}}
		\left[\sum_{s=\pm}-\frac{s \; K_1^s}{\sqrt{\frac{1}{(\gamma -i s V)^2}}}\right]^2
	,
\\
\label{eq:B1}
\rho_{\uparrow}^{z,l}=&
	g_{33}
	+
		\frac{\pi^2\alpha^2\Delta^2m/\gamma^2}{1+m g_{11}}
		\left[\sum_{s=\pm}(\gamma -i s V) K_0^s\right]^2
	+
		\frac{\pi^2\alpha^2\Delta^2m/\gamma^2}{1-m g_{11}}
		\left[\sum_{s=\pm}-\frac{s \; K_1^s}{\sqrt{\frac{1}{(\gamma -i s V)^2}}}\right]^2
	.
\end{align}
\end{subequations}

\subsection{$\bf{\hat{x}}$-polarised magnetic impurity}
\label{aapp-xpol}

Relations of the spin-LDOS on $x$ axis for upper and lower surface are calculated for the case that the magnetic impurity in direction of $\bf{\hat{x}}$ locates on the upper surface
\begin{subequations}
\begin{align}
\label{eq:B2}
\rho_{\uparrow}^{x,u}=&
	g_{11}
	-
	4m^2\pi^2\alpha^2
	\Biggl\{
		\frac{\Bigl[(V-i\gamma)K^-_0 a_+ - (V+i\gamma)K^+_0 a_-\Bigr]^2}{1-m g_{11}}
		+
		\frac{\Bigl[(V-i\gamma)K^-_1 a_+ + (V+i\gamma)K^+_1 a_-\Bigr]^2}{1+m g_{11}}
		\Biggr\}
	,
\\
\label{eq:B2}
\rho_{\downarrow}^{x,u}=&
	g_{11}
	-
	4m^2\pi^2\alpha^2
	\Biggl\{
		\frac{\Bigl[(V-i\gamma)K^-_0 a_+ - (V+i\gamma)K^+_0 a_-\Bigr]^2}{1+m g_{11}}
		+
		\frac{\Bigl[(V-i\gamma)K^-_1 a_+ + (V+i\gamma)K^+_1 a_-\Bigr]^2}{1-m g_{11}}
	\Biggr\}
	,
\\
\label{eq:B2}
\rho_{\uparrow}^{x,l}=&
	g_{33}
	-
	\frac{\pi ^2 \alpha ^2 \Delta ^2 m}{\gamma ^2}
	\Biggl\{
		\frac{\Bigl[(V-i\gamma) K_0^- - (V+i\gamma) K_0^+\Bigr]^2}{1-m g_{11}}
		+
		\frac{\Bigl[(V-i\gamma)K^-_1  + (V+i\gamma)K^+_1 \Bigr]^2}{1+m g_{11}}
	\Biggr\}
	,
\\
\label{eq:B2}
\rho_{\downarrow}^{x,l}=&
	g_{33}
	-
	\frac{\pi ^2 \alpha ^2 \Delta ^2 m}{\gamma ^2}
	\Biggl\{
		\frac{\Bigl[(V-i\gamma) K_0^- - (V+i\gamma) K_0^+\Bigr]^2}{1+m g_{11}}
		+
		\frac{\Bigl[(V-i\gamma)K^-_1  + (V+i\gamma)K^+_1 \Bigr]^2}{1-m g_{11}}
		\Biggr\}
	.
\end{align}
\end{subequations}
\end{widetext}

\end{appendix}

\end{document}